# Sr lattice clock at 1x10$^{-16}$ fractional uncertainty by remote optical evaluation with a Ca clock

A. D. Ludlow, T. Zelevinsky, G. K. Campbell, S. Blatt, M. M. Boyd, M. H. G. de Miranda, M. J. Martin, J. W. Thomsen, S. M. Foreman, and Jun Ye

*JILA, National Institute of Standards and Technology and University of Colorado, Department of Physics, University of Colorado, Boulder, CO 80309-0440, USA*

T. M. Fortier, J. E. Stalnaker, S. A. Diddams, Y. Le Coq, Z. W. Barber, N. Poli, N. D. Lemke, K. M. Beck, and C. W. Oates

*National Institute of Standards and Technology, 325 Broadway, Boulder, CO  80305, USA*

Optical atomic clocks promise timekeeping at the highest precision and accuracy, owing to their high operating frequencies.  Rigorous evaluations of these clocks require direct comparisons between them.  We have realized a high-performance remote comparison of optical clocks over km-scale urban distances, a key step for development, dissemination, and application of these optical standards. Through this remote comparison and a proper design of lattice-confined neutral atoms for clock operation, we evaluate the uncertainty of a strontium (Sr) optical lattice clock at the 1x10$^{-16}$ fractional level, surpassing the current best evaluations of cesium (Cs) primary standards.  We also report on the observation of density-dependent effects in the spin-polarized fermionic sample and discuss the current limiting effect of blackbody radiation-induced frequency shifts.

The quest to develop more accurate quantum frequency standards has produced a detailed understanding of matter-field interactions and a vast toolbox of techniques for precision measurement and quantum state control. Historically, the neutral $^{133}$Cs microwave clock transition has offered the highest realizable accuracy for state-of-the-art frequency standards (*1,2*). In recent years, interest in and performance of standards based on optical atomic transitions have grown (*3*), driven by their superior resonance quality factors (*4*). The most accurate optical clocks are presently based on single trapped ions (*5*), due to the exquisite control possible over their electronic and motional quantum states, as demonstrated in both clock and quantum information experiments (e.g. *5-8*). Although neutral atoms (e.g. *9-12*) enjoy high measurement precision from the use of large ensembles, a longstanding challenge for these optically-based systems is achieving control and measurement at similar uncertainties as for single trapped particles. We report here a systematic uncertainty evaluation for a neutral Sr optical atomic standard at the $10^{-16}$ fractional level, surpassing the best evaluations of Cs fountain primary standards. This demonstrates control of clock states for large ensembles of atoms approaching that of the best single ion systems. Precise understanding of interactions among lattice-confined atoms will allow clean preparation, control, and readout of atoms for quantum simulations (*13*). Our measurements with thousands of atoms approach the fundamental quantum noise limit, opening the possibility of spin squeezing in optical lattices to combine precision measurement and quantum optics.

Rigorous determination of clock performance can only be achieved by comparing different clocks of similar performance. Although Cs primary standards have served as the best-characterized clock references, direct comparison between optical atomic clocks has now become essential as the uncertainties in the realized, unperturbed clock frequencies are now smaller than those for the best Cs primary standards (as demonstrated in this work and the single ion clock (*5*)). These systematic uncertainties form the essential element of clock accuracy. Unfortunately, the current complexity of these high-performing optical clocks limits the availability of multiple systems in a single laboratory. Comparing remotely located clocks can circumvent this difficulty. However, traditional methods for these remote clock comparisons such as global positioning satellite links or microwave frequency networks are increasingly inadequate for transferring optical clock signals due to their insufficient stability. All-optical comparison between remote optical clocks permits measurement without compromising the clocks' high precision and ultimately will facilitate tests of fundamental physical laws (*14*) and enable long baseline gravitational measurements (*15*) or long-distance quantum entanglement networks. The optical comparison between the JILA Sr lattice clock on the University of Colorado campus and the calcium (Ca) optical clock at the NIST Boulder laboratories is accomplished remotely by a 4-km optical fiber link.

The optical link uses coherent optical carrier transfer (Fig. 1, with more technical details provided in the Supporting Online Material (*16*)). A self-referenced octave-spanning optical frequency comb at JILA is phase-locked to the Sr clock laser operating at 698 nm. A continuous wave Nd:yttrium-aluminum-garnet laser at 1064 nm is phase-locked to the same frequency comb and its light is transferred to NIST by a phase-noise-cancelled fiber link. In this way, the Sr timekeeping is phase-coherently transmitted to NIST with a measured transfer instability of $6 \times 10^{-18}/\sqrt{\tau}$ (*17*), where $\tau$ is the averaging time. A $10^{-15}$ fraction of the Sr clock frequency is 0.4 Hz. The transferred Nd:YAG laser light is then frequency-counted against another octave-spanning optical frequency comb at NIST (*18*), which is phase-stabilized to the Ca optical clock operating at 657 nm (*19*). Both comb systems have demonstrated the capability of supporting optical clocks at below the $10^{-18}$ level (*16*).

For Sr clock operation, ~4000 $^{87}$Sr (nuclear spin $I = 9/2$) atoms are laser cooled to 2.5 μK and confined in a one-dimensional (1D) optical lattice. Spectroscopic probing of the $^1S_0 - {}^3P_0$ clock transition (Fig. 2A) is performed along the strong confinement axis of the lattice, in the Lamb-Dicke regime and the resolved sideband limit (*8*). Clock interrogation is thus highly immune to Doppler and recoil effects. Using optical pumping, the atoms are (doubly) spin-polarized and occupy only the two states $m_F = \pm 9/2$ ($m_F$ = magnetic quantum number). Fig. 2B shows spectra of the π clock transitions ($\Delta m_F = 0$) with and without the optical pumping. With the spin-polarized sample under a bias magnetic field, we interrogate the isolated $m_F = \pm 9/2$ clock transitions for 80 ms, allowing Fourier-limited spectral linewidths of 10 Hz. For these conditions, quantum projection noise would limit the Sr clock stability to $7 \times 10^{-16}/\sqrt{\tau}$ (*20*). To characterize the potential signal-to-noise ratio of the present system, we excite atoms on resonance using a short, strong Rabi pulse to power and Fourier-broaden the excitation, reducing the effect of laser frequency noise. We then measure the normalized, shot-to-shot excitation fraction as we re-load the optical lattice with new samples. The excitation fluctuations are consistent with the expected quantum projection noise. However, during clock operation, the probe laser frequency noise deteriorates clock stability. High-frequency laser noise is aliased into the low-frequency measurement band because of the optical Dick effect (*21*). This effect is exacerbated by dead time (1 s) between measurements, which includes cooling the atoms, loading the lattice, polarizing the atomic sample, and determining the populations. The measured frequency noise spectrum of our probing laser (*22*) limits clock stability to $2 \times 10^{-15}/\sqrt{\tau}$.

Fig. 2C summarizes these different stability figures for the Sr lattice clock, including the measured Allan deviation between the Sr clock at JILA and the Ca clock at NIST. The Ca clock is a simple and robust system that uses freely expanding cold atoms. Like a hydrogen maser, it serves as a highly stable frequency reference, but with 100 times better stability at short times. Indeed, the stability of the Sr-Ca comparison can reach below $3 \times 10^{-16}$ after 200 s. However, the Ca clock is susceptible to long-term (>1000 s) drifts due to residual Doppler effects. To optimize evaluation of the Sr clock uncertainties, frequency measurements are thus made in 100-s time windows. To remove sensitivity to long-term Ca drifts, these 100-s windows are interleaved as a particular parameter of the Sr clock is systematically varied. Typically, the parameter is toggled between two settings for two consecutive 100-s windows, and a Sr clock frequency shift is measured between them. Many such pairs are accumulated to average to a measurement precision below $1 \times 10^{-16}$, enabling rigorous evaluations of key Sr clock frequency shifts at this level. This measurement approach is facilitated by the robustness of these optical clocks, as both systems are regularly operated on-demand for timescales of a day.

Optical confinement of the Sr atoms occurs at a "magic" wavelength where the polarizabilities of the two clock states are equal (*23-25*). The dipole polarizability can be expanded into the scalar, vector, and tensor terms with zero, linear, and quadratic dependence on the $m_F$ of the clock state, respectively (*26*). The opposite symmetries of the vector and tensor polarizabilities facilitate orthogonalization of their effects experimentally. Because of the antisymmetric $m_F$-dependence of the vector polarizability, clock stabilization to the average of the ± 9/2 transitions eliminates dependence of the clock frequency on the vector light shift. This effect would instead be observed as a change in the frequency separation between the two ± 9/2 transitions added to the Zeeman splitting from the bias magnetic field. The averaged clock transition retains dependence on the symmetric tensor polarizabilty, which simply adds a $|m_F|$-dependent offset to the dominant scalar term. Thus, for a given lattice polarization, each spin state has a well-defined magic wavelength for insensitive confinement. The lattice laser providing the atomic confinement is frequency stabilized to the optical frequency comb, which itself is locked to the Sr clock laser. At a lattice peak intensity of $I_0$ = 3 kW/cm$^2$ and at the typical operating laser frequency, we observe a 6.5(5)x10$^{-16}$ shift in the clock frequency compared to zero intensity (see Fig.3A). Combined with our previous measurement of the weak clock sensitivity to the lattice frequency (*24*), we extrapolate the magic lattice frequency to be 368,554.68(18) GHz, in agreement with previous measurements (*9, 23-25*). Although tensor shifts are estimated to be small, we clarify that this magic frequency is specified for the π-transitions from the ± 9/2 nuclear spin states.

The hyperpolarizability effect (fourth order in electric field) has been measured (*25*), and for our operating conditions is $2(2) \times 10^{-17}$.

As with the lattice vector Stark shift, the first order Zeeman shift is cancelled because of the antisymmetric linear dependence of the ±9/2 states on the magnetic field. To individually address spin states, we choose a bias field ($B_0 = 20$ µT) large enough to resolve π transitions and reduce line pulling from residual populations of other spin states but small enough to keep the spin-symmetric second order Zeeman shift small. The size of $B$ is calibrated by measuring the frequency spacing between two spin-state transitions (*26*). To determine the second-order sensitivity and the first-order insensitivity, we measure clock frequency shifts as a function of $B$. An example of one such measurement is shown in Fig. 3C. Each set of data is fit to a second-order polynomial. The fit parameters of many such sets of data are averaged, and the first-order shift is found to be $2(2) \times 10^{-17}$ (for $B_0$), consistent with zero. The second order shift is $2.3(2) \times 10^{-17}$, and the measured shift coefficient is $5.8(8) \times 10^{-8}$ T$^{-2}$, consistent with other measurements (*12*).

Although an ensemble of neutral atoms enables large signal-to-noise measurements for high precision and stability, interactions among colliding atoms can result in frequency shifts that degrade the system accuracy. For the case of lattice clocks, unity (or less)–filled sites in a 3-D lattice can keep atomic spacing to at least half an optical wavelength and thus reduce interatomic interactions (*23, 27*). For the 1-D lattice, use of identical $^{87}$Sr fermions at ultracold temperature can exploit the Pauli exclusion principle to reduce interactions (*28*) by eliminating even-wave collisions. Ground-state ($^1S_0$-$^1S_0$) and excited-state ($^3P_0$-$^3P_0$) inter-atomic potentials have been theoretically calculated (*29*), but the only experimental measurements exist in photoassociation spectroscopy of even isotopes of Sr (*30*). We have observed a density-dependent frequency shift even when the atoms are polarized to a single spin state, indicating possible p-wave interactions or the loss of indistinguishability due to inhomogeneous excitation (*16*). Varying the atomic density by changing the spin-polarized atom number in a fixed lattice environment allows us to determine this density shift precisely, using the high measurement stability of the optical comparison. This shift scales with the atomic excitation fraction and depends on the nuclear spin state. By controlling the relevant system parameters, we are able to measure and control this shift of $8.9(8) \times 10^{-16}$ (see Fig. 3B, with spin polarization to $\pm 9/2$, $\rho_0 = 1 \times 10^{11}$ cm$^{-3}$). We have also observed an excitation fraction for which the density shift of the clock transition is consistent with zero.

Table 1 gives a summary of our investigation of the Sr systematic clock shifts. The probe laser ac Stark shift, line pulling, and servo error are described in (*16*). The overall fractional systematic uncertainty is $1.5 \times 10^{-16}$, the smallest uncertainty reported for any neutral atom standard to date and represents an improvement by a factor of 6 over our previous result (*9*). This uncertainty evaluation has enabled an improved absolute frequency measurement of the Sr clock transition (*31*). The evaluation of the Sr clock is now limited by knowledge of Stark shifts of atomic energy levels induced by the room temperature blackbody radiation (BBR) (*32*), a critical issue for developing standards. The highest accuracy calculation of the BBR shift considers both bound- and continuum-state contributions, dynamical corrections to the static polarizability, and higher-order multipole contributions (*33*). At room temperature, the uncertainty in the BBR shift originating from uncertainty in the polarizability (1%) is $7 \times 10^{-17}$. Further uncertainty in the BBR shift originates from lack of control and homogeneity of the blackbody environment (at room temperature, T). By monitoring many positions on the Sr vacuum chamber, we determine the blackbody environment to $\Delta T = 1$ K (RMS, contributions from the thermal Sr oven are negligibly small), leading to a shift uncertainty of $7.5 \times 10^{-17}$. Combining the two effects yields a $1 \times 10^{-16}$ total BBR uncertainty.

To further improve the Sr accuracy, the differential static polarizability of the clock states must be known to better than 1% (with the dynamical correction contributing <5% to the total shift). This can be measured directly by enclosing the atoms in a well-characterized blackbody environment and recording the clock shift as this temperature is systematically varied (this simultaneously decreases uncertainty in the BBR environment). The technical challenge lies in the control of temperature homogeneity over various functional areas of the vacuum chamber while accommodating sufficient optical access for a variety of atomic manipulations. One possible solution is to cool and trap atoms in a standard chamber, and then transport them in a moving lattice (*34*) to a secondary chamber where an ideal, well-defined blackbody environment is established (*16*). Such an approach avoids the complexity of cryogenic operation and can be generalized to other lattice clock species. These improvements can potentially improve the BBR-related uncertainty to far below $10^{-16}$.

36. We thank X. Huang for technical help in operating the Sr clock. The research is supported by the Office of Naval Research, National Institute of Standards and Technology, National Science Foundation, and Defense Advanced Research Projects Agency. A. Ludlow acknowledges support from NSF-IGERT through the OSEP program at CU. G. Campbell acknowledges support from National Research Council. TZ present address: Columbia University, New York. JWT permanent address: Niels Bohr Institute, Copenhagen, Denmark. SMF present address: Stanford University, Palo Alto. JES present address: Oberlin College, Oberlin, OH. NP permanent address: Univ. di Firenze, Italy.

**Figure Captions:**

Figure 1: Remote optical clock comparison. The atomic reference is counted and distributed in direct analogy to traditional clocks using mechanical gears, albeit at optical frequencies. The Sr atomic clock at JILA [1, lattice trapped Sr system and a laser at 698 nm stabilized to vertically mounted high-finesse optical cavity (*22*)] serves as the atomic standard to which an optical frequency comb (2) is phase-locked. A cw Nd:YAG laser at 1064 nm (3, a non-planar ring oscillator) is phase-locked to the same frequency comb and transferred to NIST by a phase-noise-cancelled fiber link (4). The 1064 light emerges from the fiber at NIST and is phase coherent with the light originating from JILA, as symbolized by the synchronized clocks on either fiber end. The Ca optical clock at NIST (5, free space Ca atoms and a laser stabilized to a horizontally mounted optical cavity operating at 657 nm) is the atomic reference to which an optical frequency comb (6) at NIST is phase-locked. The clock comparison (7) is made by a heterodyne measurement between the NIST frequency comb and the 1064-nm transmitted light.

Figure 2: Sr clock operation. (A) Sr energy level diagram. The red and blue dashed lines indicate laser cooling transitions and the solid red line the clock transition. (B) Spectroscopic probing of the clock transition under a bias magnetic field to lift the nuclear spin state degeneracy. π transitions with (without) spin polarization optical pumping are shown in blue (green), normalized to the unpolarized ±9/2 peaks. The inset indicates the individual nuclear spin states. After spin polarization, the population resides only in the $m_F = \pm 9/2$ $^1S_0$ states. (C) Stability figures for the Sr optical clock as a function

of averaging time. Black dashed line: quantum projection noise limit. Blue filled circles: Sr-Ca remote comparison. Green open circles: in-loop stability, based on the error signal of the atom-laser servo. Red solid line: calculated stability, consisting of the free running laser stability (*22*) at short timescales, atom-laser servo operation at intermediate timescales, and the Dick effect limit at long timescales.

Figure 3: Clock shift sensitivities. Histogram of clock shift sensitivity to (A) the lattice laser intensity slightly away from the magic frequency and (B) the atomic density. (C) One of 20 data sets indicating the clock sensitivity to magnetic fields. The inset to C shows the histogram of the linear Zeeman shift evaluated at $B_0$, based on the 20 data sets.

**Table Captions:**

Table 1: Systematic frequency corrections and their associated uncertainties for the $^1S_0 - {}^3P_0$ clock transition, in units of $10^{-16}$ fractional frequency.

| Contributor | Correction ($10^{-16}$) | Uncertainty ($10^{-16}$) |
|---|---|---|
| Lattice Stark (scalar/tensor) | -6.5 | 0.5 |
| Hyperpolarizability (lattice) | -0.2 | 0.2 |
| BBR Stark | 52.1 | 1.0 |
| ac Stark (probe) | 0.2 | 0.1 |
| First order Zeeman | 0.2 | 0.2 |
| Second order Zeeman | 0.2 | 0.02 |
| Density | 8.9 | 0.8 |
| Line pulling | 0 | 0.2 |
| Servo error | 0 | 0.5 |
| Second order Doppler | 0 | <<0.01 |
| Systematic total | 54.9 | 1.5 |

**Table 1**

Table 1: Systematic frequency corrections and their associated uncertainties for the $^1S_0 - {}^3P_0$ clock transition, in units of $10^{-16}$ fractional frequency.

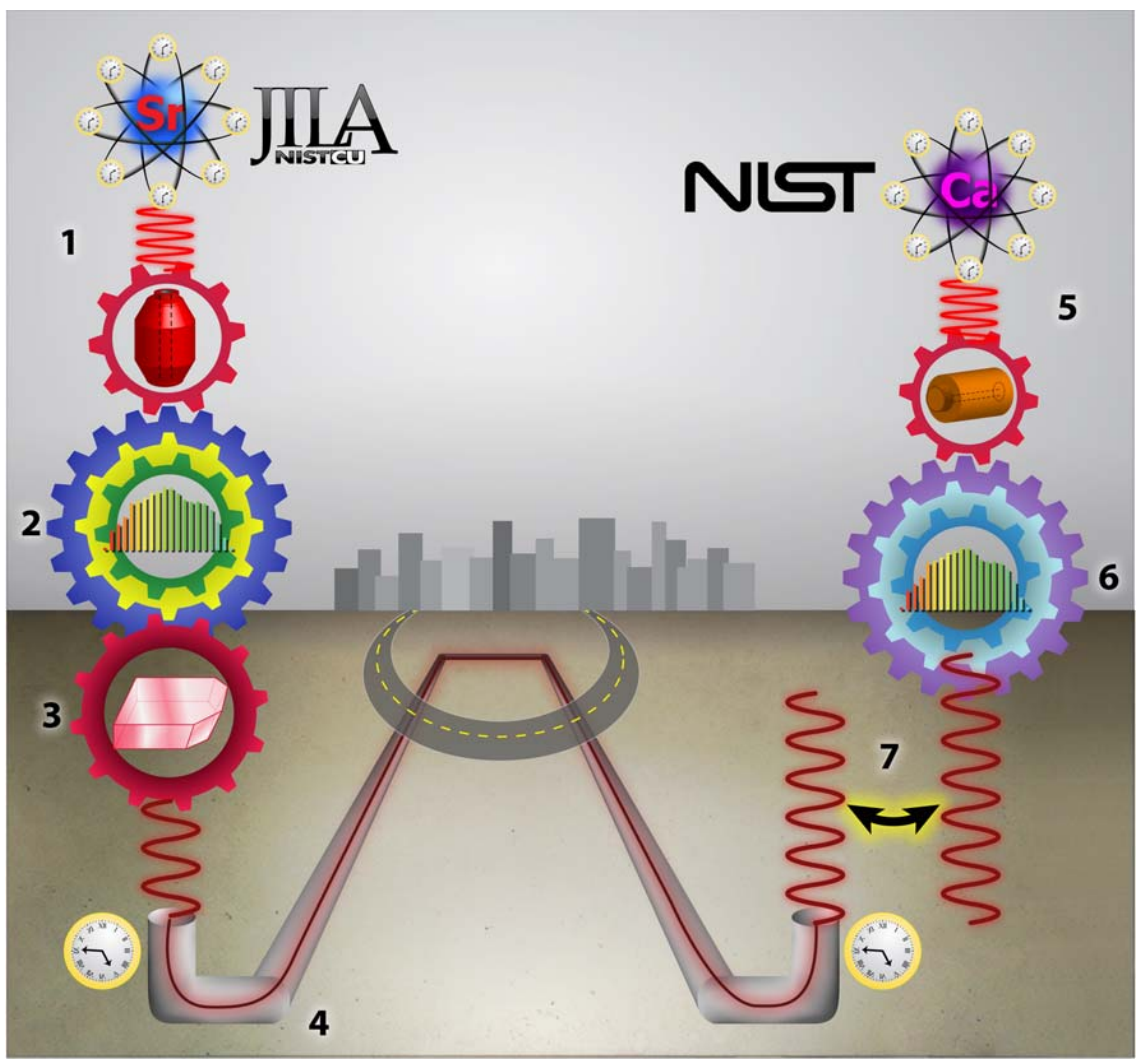

**Fig. 1**

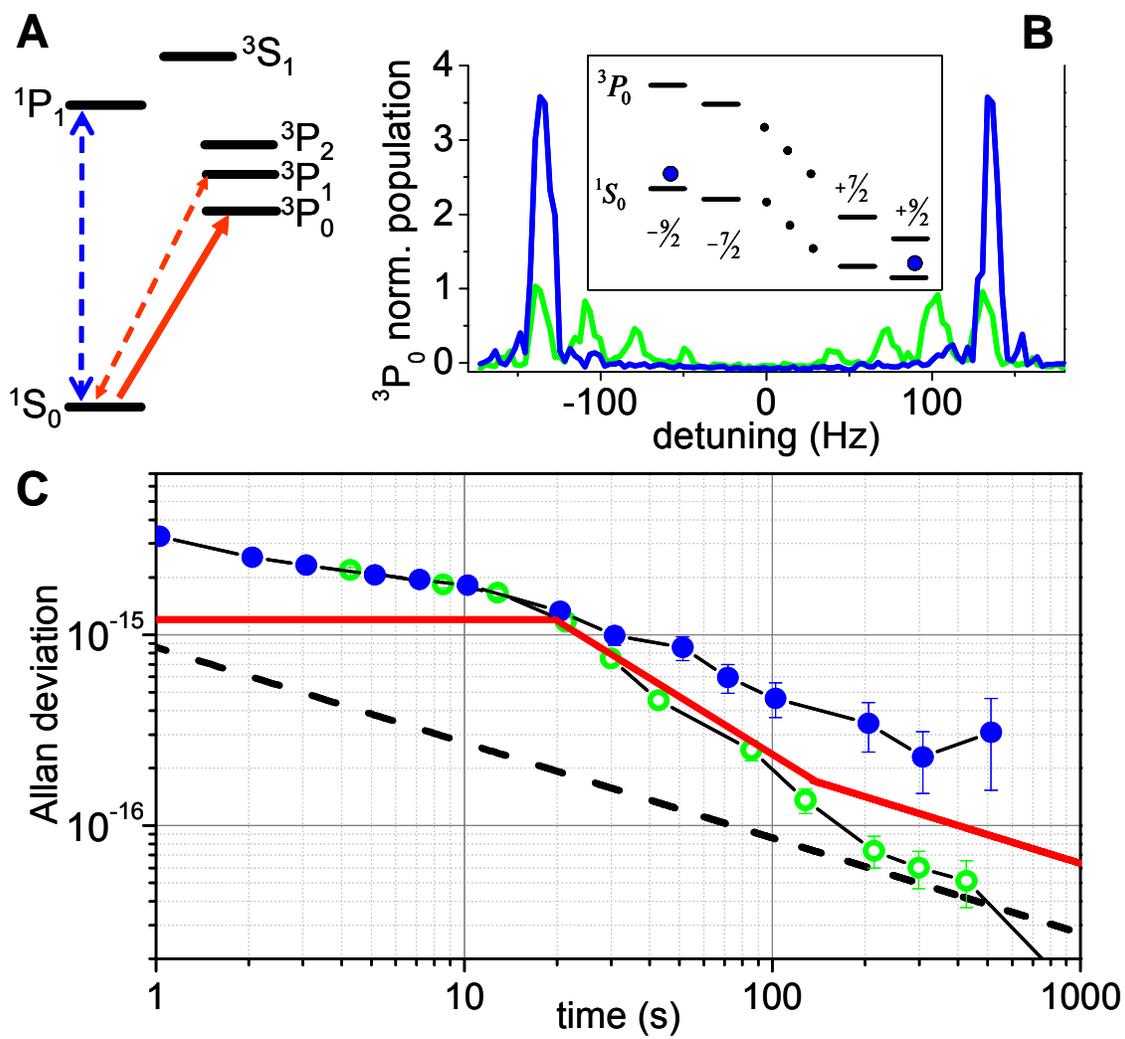

Fig. 2

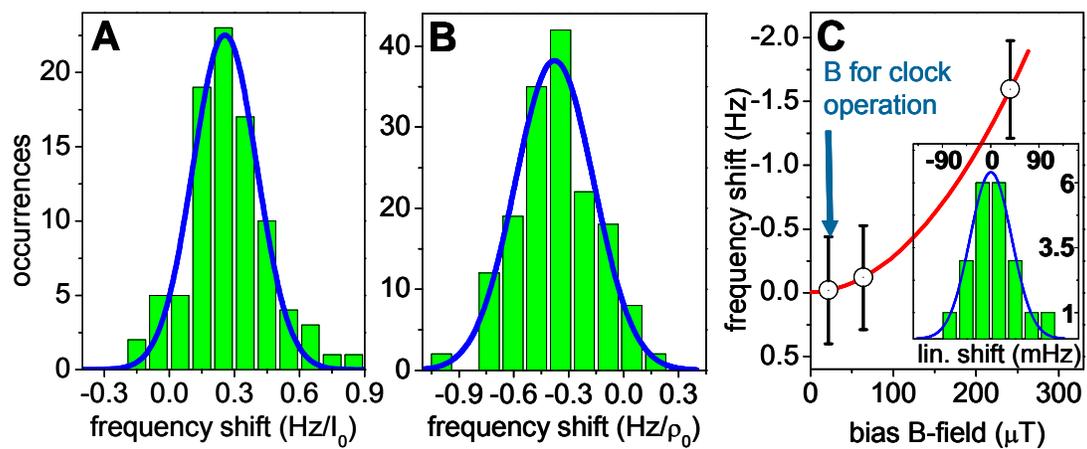

**Fig. 3**

# Supporting Online Material –
# Sr lattice clock at 1x10$^{-16}$ fractional uncertainty by remote optical evaluation with a Ca clock

A. D. Ludlow, T. Zelevinsky, G. K. Campbell, S. Blatt, M. M. Boyd, M. H. G. de Miranda, M. J. Martin, J. W. Thomsen, S. M. Foreman, and Jun Ye

*JILA, National Institute of Standards and Technology and University of Colorado, Department of Physics, University of Colorado, Boulder, CO 80309-0440, USA*

T. M. Fortier, J. E. Stalnaker, S. A. Diddams, Y. Le Coq, Z. W. Barber, N. Poli, N. D. Lemke, K. M. Beck, and C. W. Oates

*National Institute of Standards and Technology, 325 Broadway, Boulder, CO 80305, USA*

Supporting documentation is provided for our manuscript Ref. [1].

## I. ADDITIONAL EXPERIMENTAL DETAILS

To prepare the atomic sample for the Sr clock, we use narrow-line Doppler cooling techniques to reach μK temperatures. Approximately 4000 $^{87}$Sr atoms at 2.5 μK are confined in a 1-D optical standing wave generated by retroreflecting an incident laser with intensity $I_0$. The resulting longitudinal (transverse) trap frequency is 42 kHz (120 Hz). The lattice trap depth corresponds to 35 times the lattice photon recoil energy. The atom density $\rho_0$ is approximately $1\times10^{11}$ cm$^{-3}$. Further experimental details can be found elsewhere (*23,9*). Although both clock states have electronic angular momentum J=0, the nuclear spin I=9/2 permits ten spin states, all of which are populated in the ground clock state after laser cooling. We optically pump the lattice-trapped ground state population to the $m_F = \pm9/2$ states by exciting the $^1S_0$ F=9/2 - $^3P_1$ F'=7/2 transition. The efficacy of this optical pumping is shown in Fig. 1B, where spectroscopy of the π clock transitions is shown with and without optical pumping.

After clock spectroscopy, atomic population remaining in $^1S_0$ is measured by collecting $^1S_0$-$^1P_1$ fluorescence (see Fig. 2A). This nearly resonant excitation lasts a sufficiently long time to both measure the $^1S_0$ atom number and heat this population out of the trap. To measure the $^3P_0$ population, we then optically pump the $^3P_0$ population to $^3S_1$, where decay to $^1S_0$ occurs through $^3P_1$ and we again collect $^1S_0$-$^1P_1$ fluorescence. We combine these measurements to yield a normalized excitation fraction insensitive to number fluctuations of atoms loaded into the lattice. Due to occupation of a variety of longitudinal and transverse motional states, atoms in the 1-D lattice experience different Rabi frequencies from clock excitation (*8*). Inhomogeneities such as this one limit our total excitation fraction to ~80%. After probing the π-transition for $m_F = 9/2$, we re-load a new atom sample and probe the π-transition for $m_F = -9/2$. We stabilize the clock laser to these two transitions by time multiplexing two independent servos, and the digital average of the two line centers serves as the atomic reference.

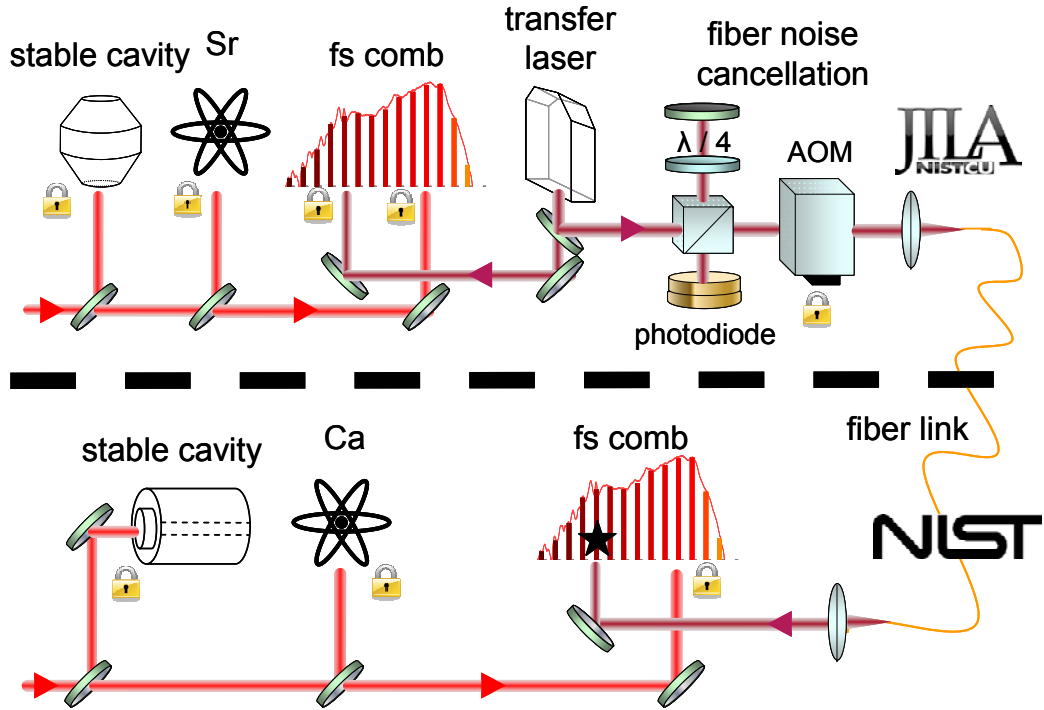

**Figure S1**

Further details on the experimental setup implemented for comparing the Sr clock to the Ca clock are shown in Figure S1. Each phase lock is pictorially represented by a pad lock. The optical local oscillator for the Sr clock is a diode laser pre-stabilized to a highly-stable, high-finesse optical cavity. The frequency of this pre-stabilized light is then locked to the Sr clock transition with a low-bandwidth servo that controls a frequency shifter between the cavity and the atomic sample. The JILA frequency comb is phase-locked to the Sr clock laser, and the Nd:YAG laser for fiber transfer is in turn phase-locked to the JILA frequency comb. The accumulated phase noise over the fiber connecting JILA and NIST is eliminated by stabilizing the round trip phase of the 1064 nm Nd:YAG light to a local copy. Strictly speaking, the Nd:YAG light received at NIST is syntonized (with exactly the same frequency, but a constant phase offset), but not synchronized, with the local light at JILA. At NIST, the NIST frequency comb is locked to the Ca clock laser, which itself is a diode laser at 657 nm pre-stabilized to a high-finesse cavity and further locked to the Ca clock transition. The measurement (location denoted by the star) is made by comparing a tooth of the NIST comb (stabilized to Ca) against the transferred Nd:YAG light (which is phase-stabilized to Sr). The techniques used to phase-lock the frequency combs and employ them in comparisons between optical standards have been verified to have a 1 s instability of 2 x $10^{-17}$ and residual uncertainty of near 1 x $10^{-19}$ [2, 3, 4, 5].

## II. ADDITIONAL DETAILS FOR SYSTEMATIC UNCERTAINTIES

Here we provide additional details of the evaluation of systematic uncertainties of the Sr optical clock summarized in Table 1 of Ref. [1].

The two clock states have different polarizabilities at the clock transition frequency. Consequently, the clock probe laser introduces a minute AC Stark shift. The small detuning of the probe laser from couplings to other motional states can lead to such shifts, but this effect is much smaller than coupling to other electronic states. Imperfect alignment between the probe and the lattice axes leads to degradation in the transverse Lamb-Dicke condition and necessitates an increased probe power to excite the clock transition. By strongly saturating the transition with excess probe intensity, we measure a non-zero shift, which is consistent with the known dynamic polarizabilities at 698 nm. By scaling to the intensity value typically used for clock operation, the shift is found to be $2(1) \times 10^{-17}$. More careful probe alignment, interrogation in a 3-D lattice, and longer probe times can further decrease the required probe power and the resulting shift.

Quantum confinement in the well-resolved-sideband limit means that Doppler effects are manifested in the sideband structure which can be isolated from the clock transition. These effects are further suppressed in the Lamb-Dicke regime where photon recoil is transferred to the optical potential. However, off-resonant, weakly excited motional transitions can pull the center frequency of the clock transition. More relevant is line-pulling originating from transitions of other spin states with residual populations after imperfect atomic polarization (typically <5%). Fortunately, many such effects largely occur in a symmetric fashion such that the line-pulling, already small, is reduced. Furthermore, the 20 μT bias *B*-field separates excitation of neighboring spin states by more than two linewidths. The overall line-pulling effect is conservatively estimated to be $<2 \times 10^{-17}$. We also note that the second-order Doppler effect is suppressed to much below $1 \times 10^{-18}$, taking into account atomic motions inside the lattice and residual motions of the lattice potential with respect to the probe beam.

Servo errors in steering the clock laser to the atomic transition can also result in frequency offsets. Our digital servo operates via standard modulation techniques by probing the half maxima of the resonance. Noise processes around the modulation frequency, round-off error, and asymmetric probe laser power spectrum are potential contributors to the servo error. In our system, we are mostly concerned with residual laser drifts and finite loop gain. We typically implement a linear feed-forward frequency compensation for drift of the stable cavity to which the clock laser is locked. This feed-forward value is estimated by measuring the laser drift relative to a Cs-calibrated hydrogen maser through the frequency comb. To overcome imperfect feed-forward compensation, we utilize two integration stages in the laser-atom feedback loop. Operating properly, this approach keeps residual drifts compensated by the primary servo integrator to a measured value of <1 mHz/s. From analysis of the overall servo signal record we conservatively estimate the servo error to be $<5\times10^{-17}$.

To improve estimation of the blackbody radiation-induced shift, the differential static polarizabilty of the clock states can be measured with a variety of approaches. One technique is to measure the static polarizability directly by observing a clock shift as a well-controlled DC electric field is applied to the atoms [6]. Another is to constrain the static polarizability value with knowledge of the dynamic polarizability. Unfortunately, at least five dipole couplings contribute to the $^3P_0$ BBR shift above the 1% level, complicating an accurate extrapolation of the dynamic polarizability to DC. Measuring the dynamic polarizability with lasers at BBR wavelengths could be effective, especially if nearby well-known atomic transitions can be used to self-calibrate the optical field strength. Finally, as discussed in the main text, one approach we propose is to cool and trap the atoms in a standard chamber, and then transport them in a moving lattice to a secondary chamber, which has a much more well-defined blackbody environment. At the highest clock accuracy level, it is important to account for the effect of the transmissivity of glass viewports for visible and infrared wavelengths on the blackbody environment. The small chamber needs accurate temperature control at the 10 mK level and would require the use of only small, far-removed optical access to introduce the optical lattice and probe beams. The effective blackbody emissivity of the interior can reach nearly unity and the total BBR environment can be known at the $1\times10^{-4}$ level, necessary for $10^{-18}$ clock uncertainty.